\pdfoutput=1
\documentclass[twocolumn]{aastex631}
\usepackage{epsf}
\usepackage{psfig}
\usepackage{color}
\usepackage{float}
\usepackage [english]{babel}
\usepackage [autostyle, english = american]{csquotes}
\pdfoutput=1 
\usepackage{graphicx} 
\usepackage[inline]{enumitem}
\usepackage{verbatim}
\usepackage{color}
\shorttitle{Ionized and H$_2$ gas in SQ}
\shortauthors{Appleton et al.}

\newcommand{\kms}{km~s$^{-1}$}

\begin{document}

\title{The interplay between ionized and warm molecular gas in the turbulent intragroup medium of Stephan's Quintet}

\author[0000-0002-7607-8766]{Philip N. Appleton}
\affiliation{Caltech/IPAC, 1200 E. California Blvd., Pasadena, CA 91125, USA}
\email{apple@ipac.caltech.edu, cosmicapple11@gmail.com}
\author[0000-0002-2421-1350]{Pierre Guillard}
\affiliation{Sorbonne Universit\'{e}, CNRS, UMR 7095, Institut d'Astrophysique de Paris, 98bis bd Arago, 75014 Paris, France}
\affiliation{Institut Universitaire de France, Minist{\`e}re de l'Enseignement Sup{\'e}rieur et de la Recherche, 1 rue Descartes, 75231 Paris Cedex 05, France} 
\author[0000-0002-9471-5423]{Ute Lisenfeld}
\affiliation{Departamento de Física Teórica y del Cosmos, Universidad de Granada, 18071 Granada, Spain}
\affiliation{Instituto Carlos I de Física Teórica y Computacional, Universidad de Granada, 18071 Granada, Spain}
\author[0000-0002-1588-6700]{C. Kevin Xu}
\affiliation{Caltech/IPAC, 1200 E. California Blvd., Pasadena, CA 91125, USA}
\author[0000-0003-1097-6042]{Francois Boulanger}
\affiliation{UPMC Universite Paris 06, {\'E}cole Normale Sup{\'e}rieure, 75005 Paris, France}
\author[0000-0003-1740-1284]{Jonathan Braine}
\affiliation{Laboratoire d'Astrophysique de Bordeaux, Univ. Bordeaux, CNRS, B18N all{\'e}e Geoffroy Saint-Hilaire, 33615 Pessac, France}
\author[0000-0001-5042-3421]{Aditya Togi}
\affiliation{Department of Physics, 749 N Comanche St, Roy F Mitte Building, Texas State University, San Marcos, TX 78666, USA}
\author[0000-0001-6217-8101]{Sarah Gallagher}
\author[0009-0001-2178-4022]{Pathayappura Aromal}
\affiliation{Department of Physics and Astronomy and Institute for Earth and Space Exploration, University of Western Ontario, 1151 Richmond St., London, ON N6A 3K7, Canada}
\author[0000-0003-2983-815X]{Bjorn Emonts}
\affiliation{National Radio Astronomy Observatory, 520 Edgemont Road, Charlottesville, VA 22903}
\author[0000-0003-0057-8892]{Loreto Barcos Muñoz}
\affiliation{National Radio Astronomy Observatory, North American ALMA Science Center, 520 Edgemont Road, Charlottesville, VA 22903}
\author[0000-0003-3343-6284]{Pierre-Alain Duc}
\affiliation{Université de Strasbourg, CNRS, Observatoire astronomique de Strasbourg (ObAS), UMR 7550, 67000 Strasbourg, France}
\author[0000-0003-0060-8887]{Benjamin Godard}
\affiliation{Laboratoire de Physique de {\'E}cole Normale Sup{\'e}rieure, Universit{\'e} de Paris, 75005 Paris, France}
\author[0000-0003-3367-3415]{George Helou}
\affiliation{Caltech/IPAC, 1200 E. California Blvd., Pasadena, CA 91125, USA}
\author[0000-0002-3471-981X]{Patrick Ogle}
\affiliation{STScI, 3700 San Martin Drive,  Baltimore, MD 21218}
\author[0000-0002-5671-6900]{Ewan O'Sullivan}
\affiliation{Center for Astrophysics $|$ Harvard \& Smithsonian, 60 Garden Street, Cambridge, MA, 02138, USA}
\author[0000-0001-8362-4094]{William T. Reach}
\affiliation{Space Science Institute, 4765 Walnut St, Suite B, Boulder, CO 80301, USA}
\author[0000-0003-0699-6083]{Tanio Diaz Santos}
\affiliation{Institute of Astrophysics, Foundation for Research and Technology-Hellas (FORTH), Heraklion, 70013, Greece.}
\affiliation{School of Sciences, European University Cyprus, Diogenes street, Engomi, 1516 Nicosia, Cyprus. }

\begin{abstract}
Using {\it JWST} MIRI spectroscopy, we explore the multi-phase intergalactic gas in the hot halo of the Stephan's Quintet compact group.  Unlike the coolest H$_2$ emission, which commonly shows double-peaked profiles corresponding to the velocity of the intruder and the intragroup gas, the ionized Neon emission-lines are often systematically shifted redwards in radial velocity away from the lowest H$_2$ peak, and into the region corresponding to gas decelerated by the collision of the intruder galaxy with intragroup gas.  The majority of the brightest clumps of ionized gas are spatially correlated with the brightest H$_2$ clumps, suggesting they are coherent structures caught up in the opposing gas flows. Gas that is progressively heated, is shifted in velocity further from the velocity of the coldest H$_2$ peaks. Larger and larger shifts in velocity are seen as one progresses from cool to warm molecular hydrogen, to lower-excitation (Neon lines) ionized and strongly shocked heated ionized gas (gas-phase Iron).  The majority of the ionized Neon in the intergalactic medium is consistent with excitation in low-velocity shocks. We suggest that this behavior is the result of gas of progressively lower densities being shocked and accelerated by the interaction of pre-existing  clouds with a hot flow created by the blast-wave from the collision between the intruder and the intragroup gas. One exception is the extragalactic star forming region, SQ-A, which shows higher-excitation ionized emission lines consistent with heating from a young starburst, consistent with previous photometry of young star clusters in the region. 

 \end{abstract}

\section{Introduction}

The Stephan's Quintet (hereafter SQ) compact group represents a remarkable laboratory for studying the way in which different gas phases can survive and interact with one another in a group-wide hot halo heated by violent collisions between the component galaxies. In this system, an intruder galaxy NGC~7318b is believed to be colliding with 
an old tidal filament in the main group at 
850-900 km $s^{-1}$, resulting in the generation of X-ray emission along an intergalactic  filament and creating enhanced emission in the wider X-ray emitting halo of the group \citep{Moles1997,Trinchieri2005,OSullivan2009}. This high-speed collision has created, in the intragroup medium, highly turbulent material comprising of a wide range of gas and plasma phases. This includes radio emitting plasma \citep{Allen1972,Xu2003}, neutral [CII] emitting gas \citep{Appleton2013}, cold and warm molecular hydrogen \citep{Guillard2012,Appleton2006,Cluver2010,Appleton2017,Emonts2025,Maeda2025,Yamamoto2026}, H$_2$O \citep{Appleton2013}, as well as diffuse ionized atomic gas \citep{Xu2003,Iglesias-Paramo2012,Konstantopoulos2014,DuartePuertas2019,DuartePuertas2021}. A remarkable result from the molecular studies is the strength and ubiquity of pure rotational lines of warm molecular hydrogen emission in the group \citep{Guillard2009,Appleton2023}, defying the simplistic expectation that molecular gas would be easily destroyed in such a high speed collision.  

In addition to the warm ($>$ 100K ) H$_2$, the long-term survival of cold dense CO-emitting clouds seen in the Stephan's Quintet IGM, (e. g. \citealt{Emonts2025}) is interesting because they would be expected to be destroyed rapidly by a wide variety of processes, including Rayleigh-Taylor and Kelvin-Helmholtz instabilities \citep{Klein1994,Zhang2017}, and mechanisms such as cloud shattering \citep{Gronke2020,Gronke2022}. Nevertheless, cold molecular gas has been seen in other hot environments, for example in filaments in the Persius Cluster \citep{Salome2006,Salome2008,Lim2012}, or  in the tails of \enquote{Jellyfish} galaxies: cluster-member galaxies which are being stripped of gas by their interaction with the hot intercluster medium \citep{Jachym2014,Jachym2017,Moretti2018,Jachym2019,Moretti2020,Moretti2023}. In these cases, somehow the clouds survive or are reformed in the turbulent flows surround the galaxies. Our study of the multi-phase mix of gas in SQ is therefore relevant in the understand the survival of gas in other halo environments. 

Unlike previous ground based studies of ionized gas in SQ which have been typically limited to large ($>$kpc) spatial scales, the {\it JWST} observations presented here provide an excellent opportunity to explore correlations and connections, at sub-arcsec scales, between the ionized and molecular gas phases particularly the low-order pure rotational transitions of H$_2$ which fall in the {\it JWST} MIRI observing window. Given the highly clumpy distribution of dust and gas seen in previous work \citep{Pontoppidan2022,Appleton2023}  multi-phase structure and correlated kinematics between gas phases on these scales seems highly likely.  More widespread, and much more extended ionized gas is also possible through the collision of very diffuse, low-density regions in the pre-shocked gas of both the intruder, and any preexisting group-wide gas. This gas is less likely to show sub-arsec-scale correlations between the warm H$_2$ and the ionized gas.

The current paper is complementary to the work done in a companion paper by our team (Guillard, et al. in prep.), which studied the survival and evolution of dense intergalactic molecular clumps in this high-speed collision. That paper, which uses the same datasets as those described here, concentrates specifically on the properties and kinematics of the warm molecular gas in SQ, and showed that the warm molecular gas has mainly survived the collision, and that some of it becomes much warmer and turbulent as it is decelerated in the reference frame of joint intruder/group system. 

In contrast, the current paper uses the same {\it JWST} MIRI-MRS spectroscopy dataset \citep{Wright2023,Wells2015,Argyriou2023} to compare the distribution and kinematics of ionized gas with that of the warm molecular gas, through the strong 0-0S(1)$\lambda17.03\mu$m pure rotational line.  While we cannot rule out very fast shocks ($>$ 700 \kms), we will present evidence that lower-velocity shocks ($>$ 50-100~\kms) are the most likely explanation for the ionization state of the [\ion{Ne}{2}]$\lambda12.81\mu$m and [\ion{Ne}{3}]$\lambda15.56\mu$m emission lines. In addition, faster shocks may explain the formation of gas-phase Iron emission (through the [\ion{Fe}{2}]$\lambda5.34\mu$m line) found in the most intense part of the shocked filament.  While most of the intragroup gas is detected in low-excitation neon lines, one exception is the region SQ-A. This is a star forming region in the far north of the main molecular filament \citep{Xu1999,Xu2025}. Here, harsher UV emission is required to explain the much higher excitation of Neon, as well as the appearance of other high-excitation lines. We will show this is consistent with photo-excitation from very young star clusters. 

The paper is organized as follows. We present in \S 2 a description of the observations and data analysis. In \S 3 we discuss the distribution and kinematics of both the warm H$_2$ and the ionized gas. In this section we present integrated maps over both the full ranges of H$_2$ and [\ion{Ne}{2}] emission, but also of selected velocity ranges, to emphasize the depth and diversity of the emission. In \S 4 we discuss previous and ongoing studies of cold molecular hydrogen through the CO lines. \S 5 describes likely excitation mechanisms for the ionized gas. This section also discusses how cold molecular gas may survive in such a harsh collisional environment, and compares our results with that of ram-pressure stripped tails in Jellyfish galaxies. \S 7 presents the conclusions. The Appendix describes how we use a comparison of the observed emission lines of Neon and Sulphur, with both shock and starburst models, to attempt to understand the excitation of the ionized gas throughout the group.

The systemic heliocentric radial velocity of the intruder galaxy, NGC~7318b, based on optical emission lines is 5774~\kms. This compares with the much higher radial velocities of the other nearby group members, NGC~7317 (6599~\kms), NGC~7318a (6663~\kms), and NGC~7319 (6747~\kms)\footnote{All radial velocities from the NASA Extragalactic Database (NED)}. At an assumed distance of 94 Mpc for Stephan's Quintet \citep{Appleton2023},  1~arcsec corresponds to 456 pc.

\section{{\it JWST} Observations and Data Analysis}\label{sec:observations}
Observations of Stephan's Quintet were made with the Mid-Infrared Instrument (MIRI) Medium Resolution Spectrometer (MRS) as part of the GO2 cycle (proposal ID: 3445, PI: P. Appleton) on 11/06/2023 and 11/07/2023.  
Spectra at each position were obtained over the full range of the MIRI MRS at each position. The full spectrum observable by the MIRI MRS (4.9 to 27.9$\mu$m) is built up by observing in three grating setting sequentially (Low, Medium and High), each requiring a separate integration. During each of the three grating observations, data is acquired simultaneously in the four different (Channel 1 through 4) integral field units (IFUs). Channel 1, 2, 3 and 4, cover wavelength ranges 4.9-7.65, 7.51-11.7, 11.5-17.98 and 17.7-27.9 $\mu$m respectively. The average detector pixels sizes are 0.196, 0.196, 0.245 and 0.273 respectively for Channel 1 through 4. The four IFUs have different (mainly cospatial) footprints on the sky, with Channel 1 having the smallest field-of-view (3.2 x 3.7 arcsecs$^2$) and Channel 4 the largest (6.6 x 7.7 arcsec$^2$).

\begin{figure*}
\includegraphics[width=0.99\textwidth]{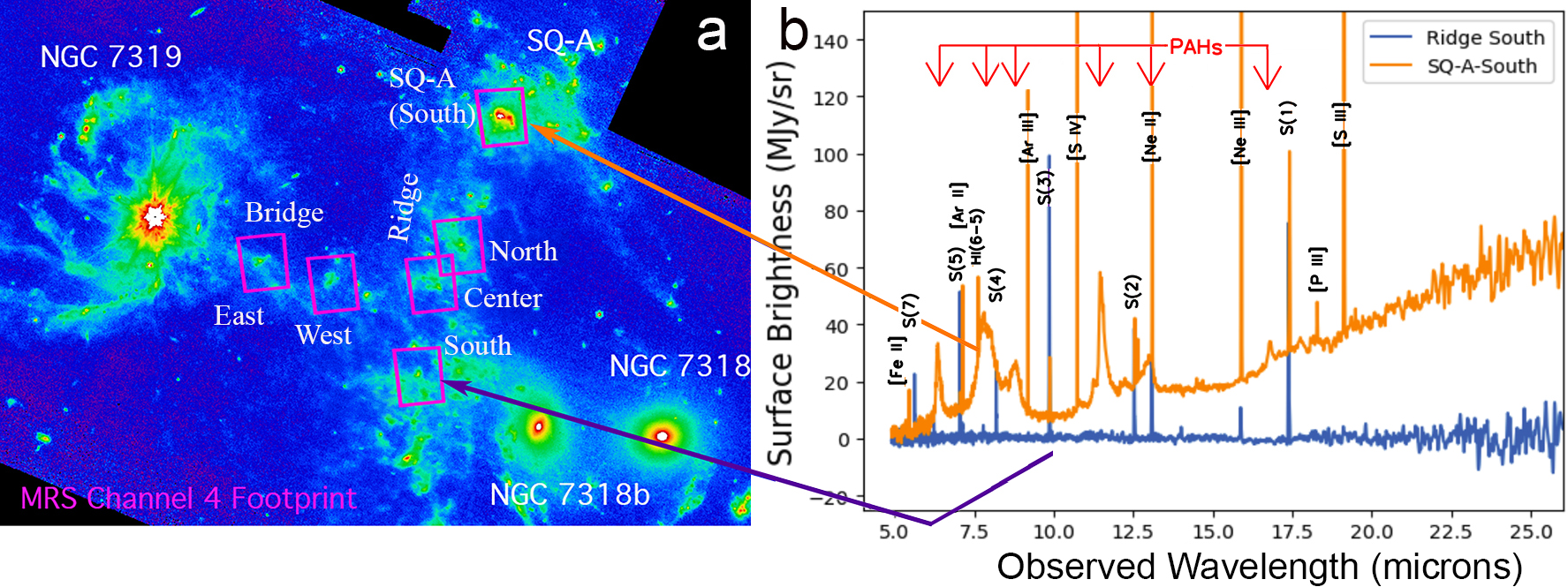}
\caption{(a) Overlay of the MIRI-MRS (Channel 4) footprints of the 6 science observations superimposed on a false-color image of the MIRI F1000W image of the inner regions of SQ. The observations target intergalactic gas along the main N/S ridge (labeled \enquote{Ridge South}, \enquote{Ridge Center}, \enquote{Ridge North} and \enquote{SQ-A (South)}, and two bright clumps of H$_2$ emission along the bridge \enquote{Bridge East} and \enquote{Bridge West}, (b) full spectra extracted from a 1 arcsec diameter aperture centered on \enquote{SQ-A-South} (orange) and \enquote{Ridge South} (blue). Important emission lines, including the rotational 0-0 S(1)-S(7) lines, and PAH bands are identified.}
\label{fig:MRSpointings}
\end{figure*}
Six MRS pointings where made within the inner Stephan's Quintet group sampling the warm H$_2$ emission seen in the intragroup medium previously by \citet{Cluver2010}. One additional position was observed outside the main body of the group to provide a background reference observation. A detailed description of the observing parameter (dither pattern, observation sequence, readout mode) is presented in a companion paper Guillard et al. (in prep.). Figure~\ref{fig:MRSpointings}a shows the placement of the MRS observations (channel 4 field-of-view 6.6 x 7.7 arcsec$^2$) superimposed on an archival 10$\mu$m NIRCam image \citep{Pontoppidan2022} of the group. Five of the regions (Ridge Center, Ridge South, Bridge East, Bridge West and SQ-A (South) were deliberately centered on previously studied areas for which either HST UV observations or high-resolution ALMA CO observations were available \citep{Guillard2022,Appleton2023}. The Ridge North region was positioned on a bright warm H$_2$ features seen in the NIRCam F770W and F1000W images of \citep{Pontoppidan2022}. Most of the observing positions were observed for a total integration time of 1.75 hr in 4 dithers per pointing, with the exception of the fainter Bridge West and the reference \enquote{off} position) which were observed for 2.63 hr each. 

The Level~0 data (pre-processed raw telemetry data) was processed through Level~1 (detector-level corrections), Level~2 (calibration of slope images), to Level~3 (science-ready IFU data cubes and 1-d spectra) using a modified version of the science pipeline to take advantage of the subtraction, from all positions, of the reference observation. The processing also included some editing of known bad pixels. The current paper uses as its starting point, flux and wavelength calibrated data cubes. We refer to Guillard et al. in prep. for a more complete description of the full processing. 

Data analysis of the IFU cubes was performed with custom-built Python software which leveraged heavily from standard software packages such as SCIPY \citep{scipy2001} and ASTROPY \citep{astropy2013,astropy2018,astropy2022}. 
Visualization and spectral extraction of these data was performed with SAOimage/DS9 (\citep{Joye2003}, and {\it JWST} custom packages, e. g. Jdaviz/Cubeviz \citep{jdadf2024}. Where individual emission lines were investigated, local low-order polynomial baselines (up to second order) were fitted to each target emission line over the whole IFU subcube. Integrated maps (Moment0 maps) of the [\ion{Ne}{2}]$\lambda12.81\mu$m and 0-0S(1) H$_2$$\lambda17.03\mu$m lines were created by summing the emission in the lines over the range in which emission is detected. We also present position-velocity maps 
of several of the observed fields using the SPECTRACUBE \citep{Ginsburg2015} and ASTROPY/PATH python packages \citep{ astropy2022}.  
In a future, more detailed paper, we will present higher moment maps and individual channel maps and well an a full description of the spatio-kinematic structures in all the observed regions.  

As an example of the full wavelength coverage of MIRI/MRS, we show  two spectra taken from two very different regions in Figure~\ref{fig:MRSpointings}b. Most of the regions observed,  including \enquote{Ridge South} (blue spectrum), show a very weak continuum and spectra dominated by warm rotational H$_2$ lines and faint [\ion{Ne}{2}] and [\ion{Ne}{3}] lines.  In contrast, the spectrum of SQ-A (orange spectrum) contains a rising IR continuum, strong PAH features and many ionization lines.  

Ionized gas emission line flux ratios, presented in the Appendix, were based on spectral extractions from circular apertures obtained in regions across all the target observations.  Only a subset of the spectra and apertures are shown in \S 3 for brevity. Line fluxes were obtained by fitting Gaussian profiles to the lines, and then converting the results to W m$^{-2}$ for each aperture, and the line ratios calculated. A more complete description of the apertures and fitting process will be presented in a forthcoming paper. 


\section{Distribution and Kinematics of the warm H$_2$ and Ionized Gas}

\begin{figure*}[ht]
\includegraphics[width=0.95\textwidth]{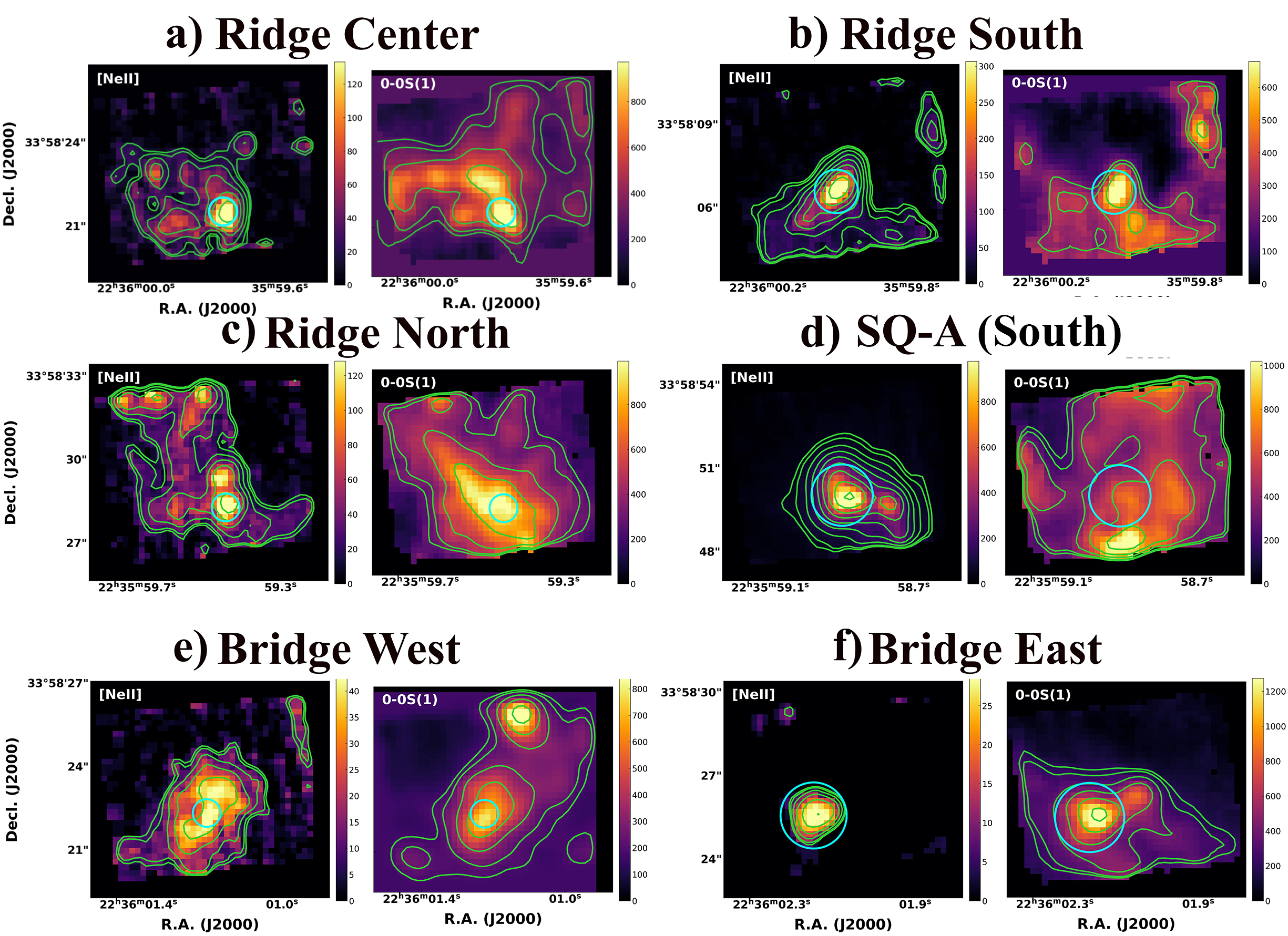}
\caption{Integrated line emission maps (Moment 0) for the six observed regions (a through f). Each panel shows the [\ion{Ne}{2}]$\lambda12.81\mu$m and 0-0~S(1)H$_2\lambda17.03\mu$m images. The cyan circles on the images show the size and positions of the extracted spectra shown in Fig.~\ref{fig:ionspectra}. The contours in all panels start at approximated 2.5 sigma and increase in multiples of 1.22. The contours in panels a), b) and c)  start at 18 MJyr sr$^{-1}$ (260 MJyr sr$^{-1}$), in panel d) at 50 (200 MJy sr$^{-1}$)  and in panel f) at 5 (200 MJy sr$^{-1}$) for [\ion{Ne}{2}] and for  0-0~S(1) line respectively.  Note that many of the H$_2$ images have quite extended emission over much of the IFU, sometimes extending to the edges of the field-of-view (as in SQ-A (South).} 
\label{fig:ionsmom0}
\end{figure*}

In Figure~\ref{fig:ionsmom0}a-f we make a comparison between the integrated (moment 0) [\ion{Ne}{2}]$\lambda$12.81$\mu$m ionized gas and the 0-0S(1) H$_2$ line for each of the six target regions. While several of the regions (Ridge Center, Ridge South and Bridge West) show broadly similar [\ion{Ne}{2}] and warm H$_2$ distributions, others, such as Ridge North, SQ-A (South) and Bridge West, show different integrated ionized and molecular gas distributions. SQ-A (South) is particularly striking, in that the main very strong ionized gas structure (a double extended structure) shows almost no obvious comparison with the warm H$_2$ which is distributed widely across the IFU field of view, and contains many different kinematic components. In Bridge East (Figure~\ref{fig:ionsmom0}f) which lies in the bridge region close to NGC~7319, the ionized gas peaks in a similar place to the peak in the warm H$_2$, but is more compact and has a radial velocity which places it between two warm H$_2$ features. Indeed it is often the case that the molecular hydrogen 
has multiple velocity components along the line of sight, whereas the ionized gas has fewer components. 

One of the regions, Ridge Center (Figure~\ref{fig:ionsmom0}a) lies near the center of the main H$_2$ filament, and shows a similar head-tail morphology in both ionized gas and H$_2$. It's  molecular structure was discussed in detail in \citet{Appleton2023} in the context of the cloud shattering of a dense cold head into a tail of entrained and phase-mixed material. 

\subsection{A broad look at the three main velocity regimes in the warm H$_2$ and Ionized gas indicators}

Although a full discussion of the spatio-kinematics of the warm H$_2$ and ionized gas is deferred to a subsequent paper where we will compare these data with new high-resolution CO observations of the cold molecular gas, we will provide here a brief overview of the distribution of the gas integrated over three main velocity regimes. For the main filament and SQ-A, this includes the gas loosely associated with the intruder (5840-6150 \kms heliocentric), the gas at intermediate velocities (6150-6650 \kms), and the gas mainly associated with the main velocity of the group (6650-7050). For the two bridge regions, we modify these ranges slightly to take into account the different kinematics seen in the bridge. Figure~\ref{fig:channels} presents maps of the integrated emission applied over three main heliocentrcic velocity ranges described above.  The figure includes all six targeted regions in Figure~\ref{fig:channels}a-f.  Each sub-figure shows a horizontal panel of the integrated emission in the 0-0S(1) line and {NeII] line respectively for the three velocity regimes. For the two bridge regions in  Figure~\ref{fig:channels}e and f, the velocity ranges for the low, intermediate and high velocities are 5950-6300, 6300-6550 and 6550-7050 \kms~respectively. The reason for this restricted range of velocities compared with the other regions will become apparent when we discuss the spectra in the next subsection.  

Figure~\ref{fig:channels}a and b) (Ridge Center and Ridge South) lie in the main N/S molecular filament, and have some similar properties when viewed in the three velocity regimes. The warm H$_2$ (upper panels) have most of the warm emission at the velocity of the intruder and this fades away as we proceed to higher velocities. In contrast, the brightest [\ion{Ne}{2}] emission is seen at intermediate velocities. This will be discussed further in the next subsection. In both the warm H$_2$ maps, a compact clump is seen extending over all velocities.

Figure~\ref{fig:channels}c and d (Ridge North and SQ-A) lie in the upper part of the main H$_2$ ridge. Both show fainter emission at low and intermediate velocities in the warm H$_2$ line, with most of the emission occurring in the higher velocity channels (above 6650 \kms heliocentric). Unusually for Ridge North,  the [\ion{Ne}{2}] emission is brightest in filamentary clumps at the intruder velocity with only faint H$_2$ counterparts at those velocities, whereas, the warm H$_2$ high-velocity shows similar filamentary structure, but the brighter clumps do not coincide between the ionized and warm H$_2$ gas. There is some faint [\ion{Ne}{2}] emitting gas associated with intermediate velocities where the emission is quite extended. 

For SQ-A, the bright ionized gas is very evident in the starburst region in the lower panel of Figure~\ref{fig:channels}d, in the  middle and upper velocity maps. This gas is associated with the starburst (as was discussed by \citet{Appleton2023}). There is a faint counterpart to this gas in the warm H$_2$ maps (upper panel) but much of the H$_2$ is dominated by other structures. 

The bridge regions are shown in Figure~\ref{fig:channels}e and f. The overall velocity spread in these features is narrower than the main field, and so we have made maps which cover a narrower range of velocities than in the previous cases. Bridge West, Figure~\ref{fig:channels}e, shows narrow linear structure in both the H$_2$ and [\ion{Ne}{2}] maps at intermediate velocities, whereas at low and high velocities the shape of the structure is more clumpy. For Bridge East, Figure~\ref{fig:channels}f, the [\ion{Ne}{2}] emission is quite compact over all the velocity maps. In the warm H$_2$ maps, although the same structure is seen, a high-velocity clump is seen to the north-west which has no counterpart in ionized gas. Again, it is generally the case for both bridge regions that the ionized gas is quite strong at intermediate velocities.   

\subsection{New insights from a study of the  MIRI spectra}
The MIRI spectroscopy provides for the first time the ability to compare the kinematics of the ionized gas with that of the warm molecular gas. In Figure~\ref{fig:ionspectra}a-f, we present representative spectra from each of the six fields. The spectra were extracted from the cyan circular regions shown in Figure~\ref{fig:ionsmom0}. The observations show several remarkable results. 

Firstly, as described in Guillard et al. in prep., with the exception of SQ-A (South), the warm molecular hydrogen (red spectra) show a double peaked profile, which is particularly pronounced along the main H$_2$ ridge (Ridge Center, South and North shown in Figure~\ref{fig:ionspectra}a, b and c). In these cases the main H$_2$ peaks correspond to gas at the most extreme velocities, corresponding to that of the intruder (around 6000 km/s), and that of the approximate barycenter of the main group (6700 km/s). As one proceeds north along the ridge, the most extreme velocity seen in the double peaked spectra increases to from 6750 to 6900 \kms.  Even the spectrum of SQ-A (Figure~\ref{fig:ionspectra}d), which shows multiple components, exhibits H$_2$ emission at these higher velocities.   In Figure~\ref{fig:ionspectra}e and f (regions Bridge West and East)  similar double-peak profiles are observed in the molecular gas, but separated by a smaller velocity differences than the main ridge. In all these cases, warm molecular gas is also seen more weakly at intermediate velocities between the main peaks. For example in Figure~\ref{fig:ionspectra}a, gas intermediate between the peaks is seen as a clear red wing on the lowest velocity peak and a blue wing on the highest velocity peak.

\begin{figure*}[p]
\centering
\includegraphics[width=0.95\textwidth]{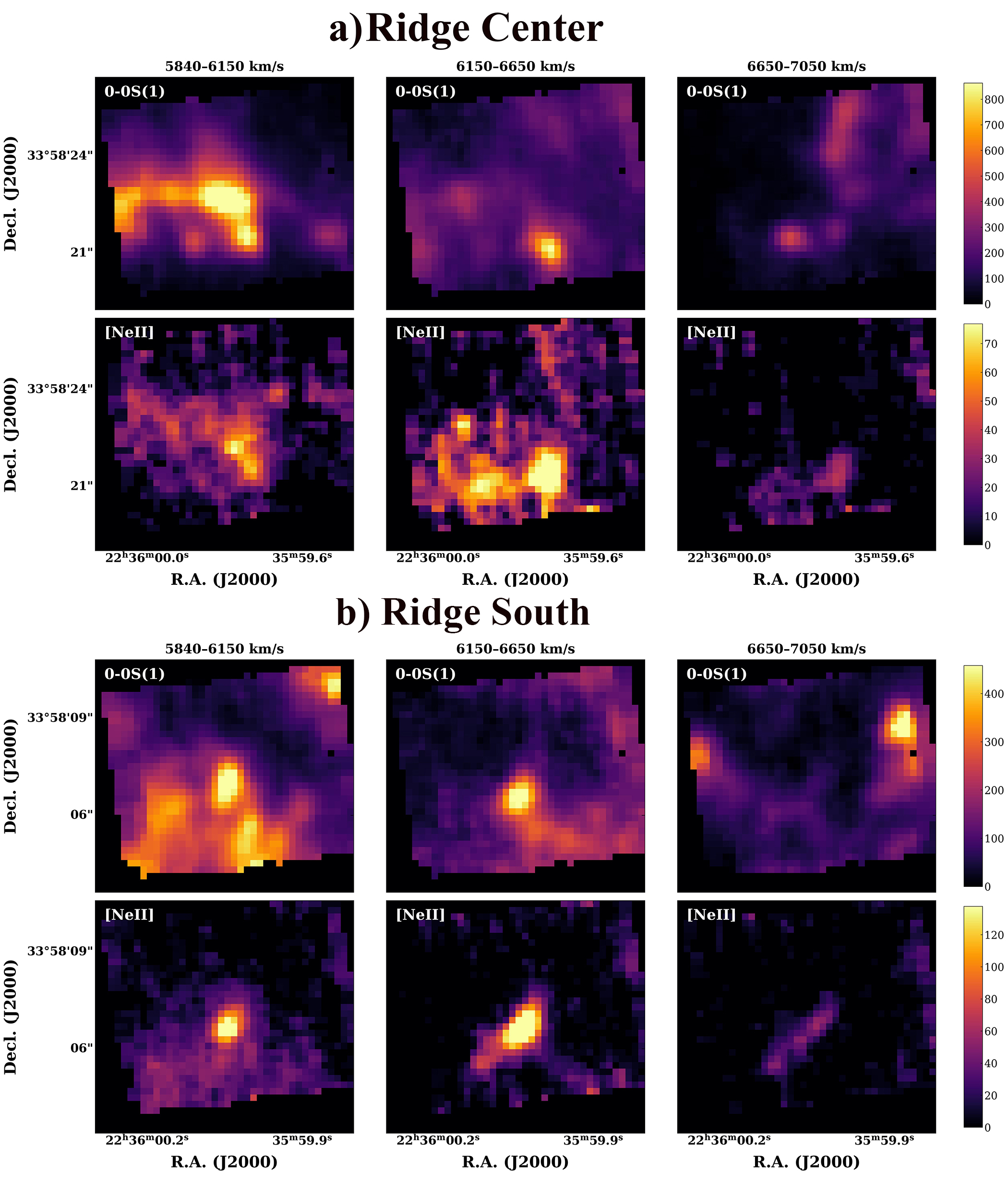}
\end{figure*}

\begin{figure*}[p]
\centering
\includegraphics[width=0.95\textwidth]{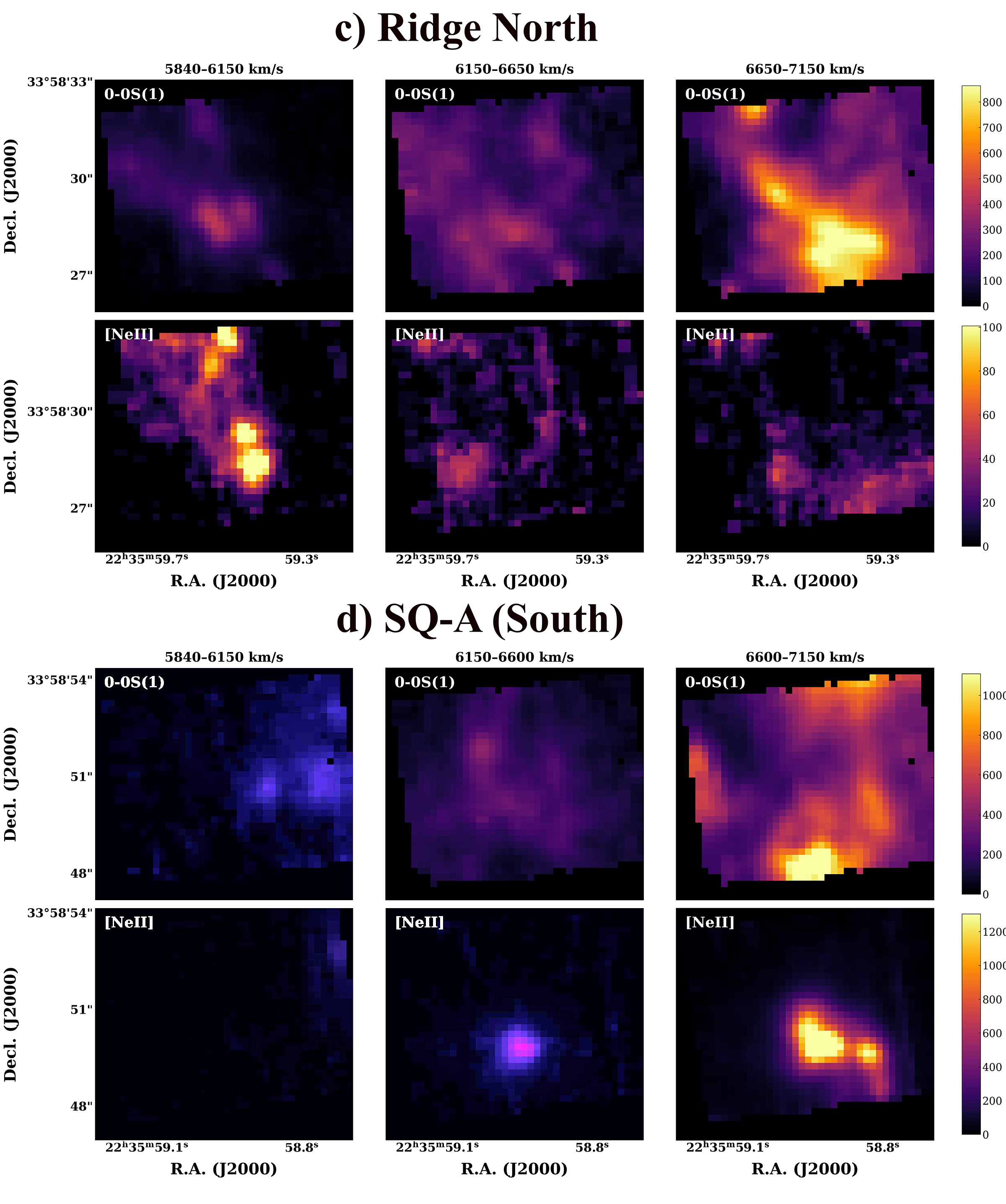}
\end{figure*}

\begin{figure*}[p]
\centering
\includegraphics[width=0.95\textwidth]{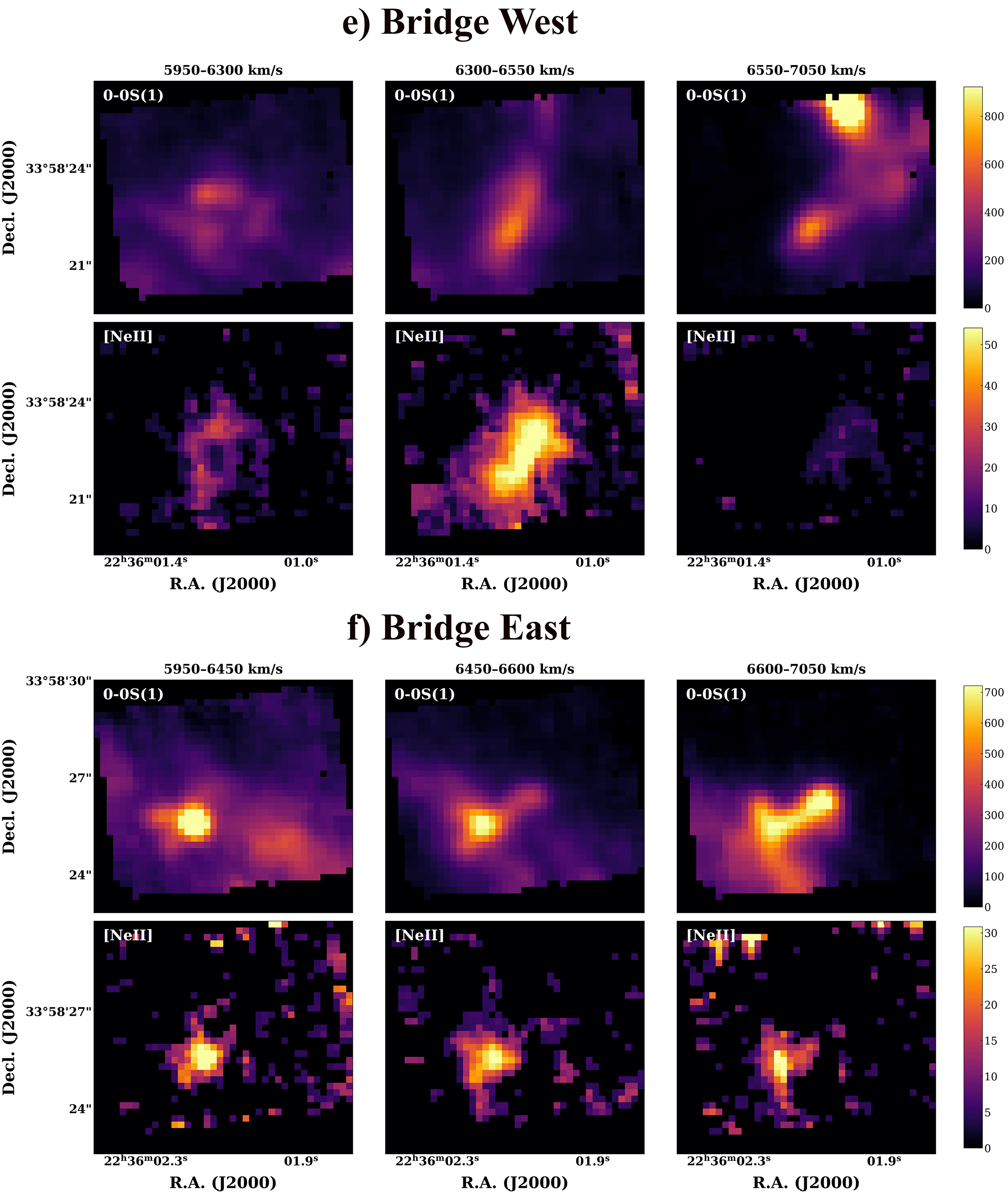}

\caption{Integrated emission for the 0-0 S(1) line of molecular hydrogen and [\ion{Ne}{2}]  of the three main velocity
regimes in Stephan's Quintet; a) Ridge Center, b) Ridge South, c) Ridge North, d) SQ-A (South), e) Bridge West, and f) Bridge East. The regions covered are shown in Figure~\ref{fig:MRSpointings}. The scale is in MJy sr$^{-1}$. The velocity range over which the integrated emission is evaluated is given at he top of each figure. }
\label{fig:channels}
\end{figure*}

\begin{figure*}
\includegraphics[width=0.95\textwidth]{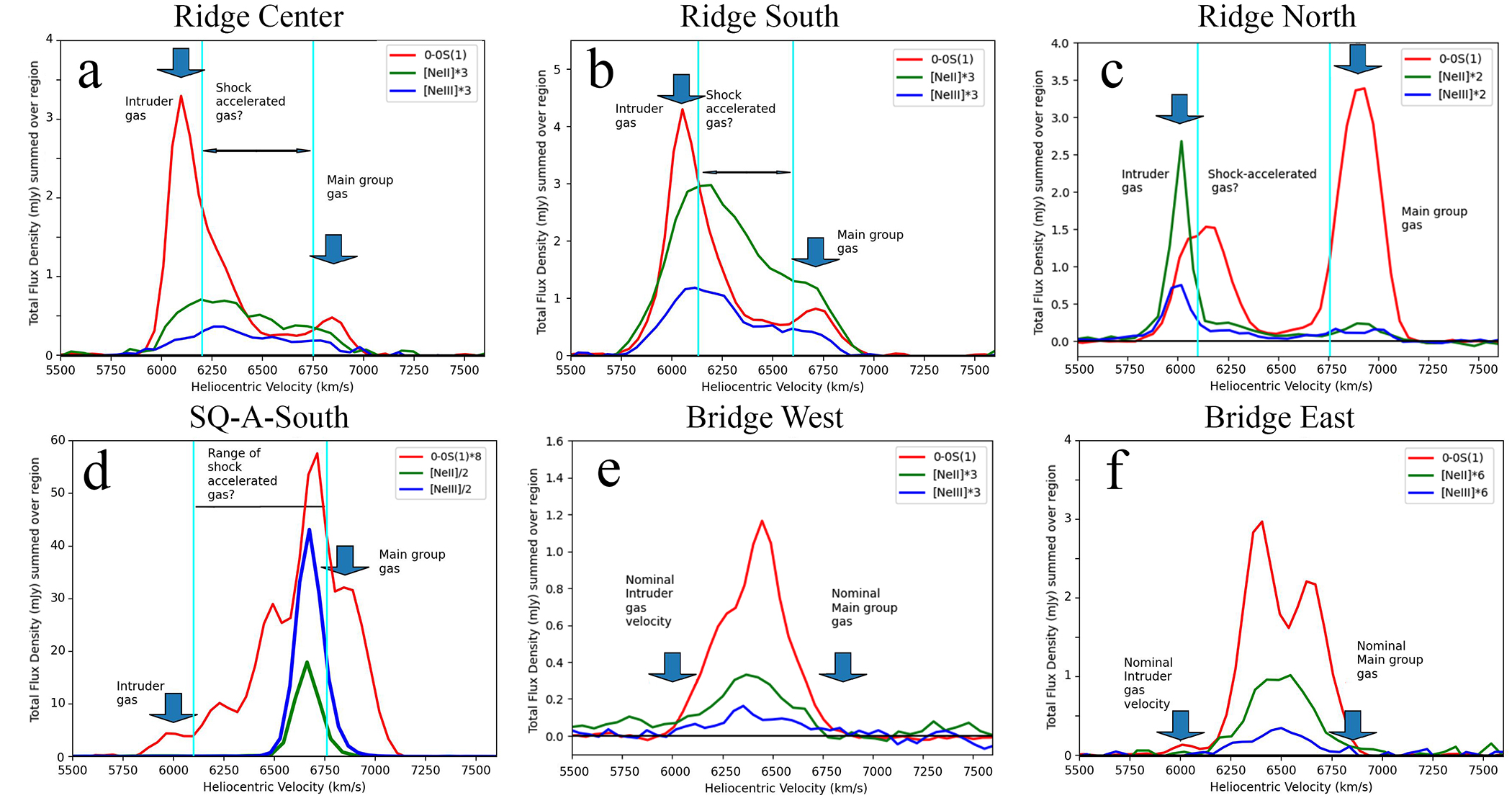}
\caption{Representative spectra extracted from each of the 6 target regions (see circles in Figure~\ref{fig:MRSpointings} for extracted apertures). a) corresponds to Ridge Center, b) Ridge South, c) Ridge North, d) SQ-A (South), e) Bridge West and f) Bridge East. The red line shows the spectrum of the 0-0 S(1)17.03$\mu$m H$_2$ line, whereas green and blue show the spectrum of the [\ion{Ne}{2}] and [\ion{Ne}{3}] lines. Note that in SQ-A (South), the [\ion{Ne}{3}] is significantly brighter than [\ion{Ne}{2}], unlike the other regions. Vertical arrows show the likely positions of gas associated with the intruder's velocity (5800-6000 km/s) and gas that may be associated with the main intragroup medium (6600-6800 km/s). Vertical cyan lines show suspected velocities where gas may have been accelerated or shock-heated in the collision.} 
\label{fig:ionspectra}
\end{figure*}

Secondly, in contrast to warm H$_2$, the kinematics of the ionized Neon (both [\ion{Ne}{2}] (green spectra) and [\ion{Ne}{3}] (blue spectra)) tend to lie at intermediate velocities between the two peaks of the warm H$_2$. For example in Ridge Center and Ridge South (Figure~\ref{fig:ionspectra}a and b), the ionized gas peaks are shifted redward of the lowest velocity (6100 \kms) H$_2$ peak by several hundred \kms, and shifted blueward of the highest velocity H$_2$ peak (6800 \kms) by a similar amount. If we interpret the double peaked H$_2$ emission as associated with gas at the velocity of the intruder and that of the main group respectively, then this suggests that the ionized gas has been decelerated (relative to the intruder) by the interaction between the two components in a similar way that we see the red and blue wings on the main H$_2$ peak in the same spectrum. In the Bridge East and West regions Figure~\ref{fig:ionspectra}e and f, the ionized Neon piles up at the velocities centered between the two warm H$_2$ velocity peaks. Only in the case of Ridge North (Figure~\ref{fig:ionspectra}c), do we see ionized gas peaking near the warm redward (6000 \kms) H$_2$ peak.  In this case the main Neon peak is narrow and blue-shifted compared with the broad (likely double-peaked) lower-velocity H$_2$ component. This rather different behavior of the [\ion{Ne}{2}] gas compared with the other regions can be seen in the maps of Figure~\ref{fig:channels}. However, even in this position, there is evidence of a long blueward tail of ionized gas that extends between the two main peaks of the warm H$_2$. We note that the [\ion{Ne}{2}] and [\ion{Ne}{3}] line profiles are often very similar, suggesting they are associated with the same gas components and share similar kinematics. 

\begin{figure*}
\includegraphics[width=0.99\textwidth]{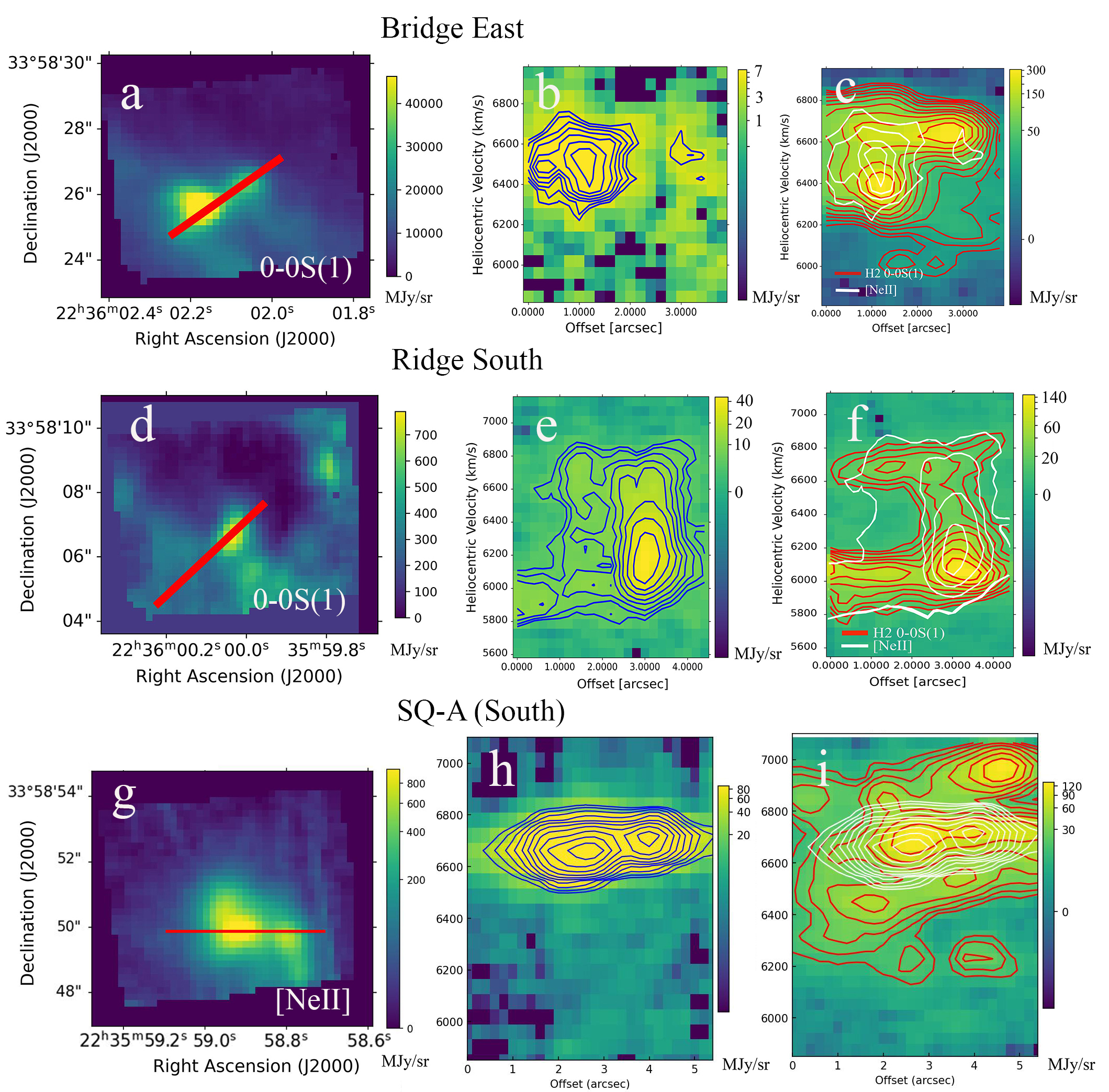}
\caption{Position Velocity (PV) Diagrams: a) Moment0 0-0 S(1) image of Bridge East, red line shows slice; b)  PV diagram for [\ion{Ne}{2}] for Bridge East (blue contours), c) PV diagram for 0-0S(1) H$_2$ line  for Bridge East (red contours) with [\ion{Ne}{2}] contours of b) superimposed (white contours); d)-f) the same as a)-c) but for Ridge South; g) Moment0 [\ion{Ne}{2}] image of SQ-A (South) with red slice marked; h)-i) same as a)-c) but based on SQ-A (South).  For g) we chose [\ion{Ne}{2}] rather than 0-0S(1) because it shows better the main ionized gas region associated with the starburst (see Figure~\ref{fig:ionsmom0}d).} 
\label{fig:PVdia}
\end{figure*}

To further explore the relationship between the ionized gas and the warm H$_2$, Figure~\ref{fig:PVdia} shows two representative position-velocity (PV) slices in cases where the ionized gas lies at intermediate velocities between the twin-peaks of the warm H$_2$.  We add a third PV diagram of the starburst region in SQ-A (South). Despite the H$_2$ emission for SQ-A showing considerable complexity, the kinematics of the [\ion{Ne}{2}] relative to the H$_2$ emission show some similarities with the other two cases. 

Figure~\ref{fig:PVdia}a and Figure~\ref{fig:PVdia}d present maps of the Moment0 (integrated) maps of the 0-0S(1) H$_2$ line for Bridge East and Ridge South respectively, and subsequent panels show corresponding position-velocity slices taken through the data cubes for the ionized [\ion{Ne}{2}] (blue contours in panels 
Figure~\ref{fig:PVdia}b and Figure~\ref{fig:PVdia}e) and 0-0 S(1) H$_2$ (red contours in panels Figure~\ref{fig:PVdia}c and Figure~\ref{fig:PVdia}f) for both regions. To aid in seeing the relationship between the [\ion{Ne}{2}] and H$_2$ PV diagrams, some of the [\ion{Ne}{2}] PV contours are superimposed in panels Figure~\ref{fig:PVdia}c and Figure~\ref{fig:PVdia}f  as additional white contours. 

Figure~\ref{fig:PVdia}g-i presents, for completeness, a PV diagram for SQ-A (South).  To emphasize the main HII region associated with the starburst in SQ-A, Figure~\ref{fig:PVdia}g shows the position of the PV slice on the [\ion{Ne}{2}] image because it bests defines the ionized gas regions. As with the previous PV diagram examples, Figure~\ref{fig:PVdia}h and Figure~\ref{fig:PVdia}i, show the actual PV diagrams for [\ion{Ne}{2}] and 0-0S(1) respectively. 

The PV diagrams for Bridge East, along a line connecting the two brightest H$_2$ blobs (Figure~\ref{fig:PVdia}d and Figure~\ref{fig:PVdia}e), confirms that the velocity distribution of the [\ion{Ne}{2}] emission lies between the two peaks of the H$_2$ emission, and is centered around 6500 km/s. The brightest [\ion{Ne}{2}] feature is spatially coincident with the brightest (low-velocity) H$_2$ feature. Similarly, fainter intermediate velocity [\ion{Ne}{2}] emission is also seen spatially near the position of the brightest part of the high-velocity H$_2$ feature.

The PV diagram for Ridge South also supports a spatial association between the brightest clumps of ionized and molecular hydrogen. Taking a slice along a position angle of the brightest elongated peak in the Moment0 map (Figure~\ref{fig:PVdia}d) we see in the [\ion{Ne}{2}] (Figure~\ref{fig:PVdia}f) gas showing an elongated structure in velocity space centered at 6200 km/s and extending to higher intermediate velocities. Effectively the [\ion{Ne}{2}] emission fills in velocities between the two extremes of the brightest H$_2$ emission (Figure~\ref{fig:PVdia}f). As we have previously shown with just the spectra, the brightest [\ion{Ne}{2}] emission is clearly offset to higher velocities (6200 km/s) from the brightest low-velocity H$_2$ feature at 6000 km/s.  The offset is easily seen in the white contours in Figure~\ref{fig:PVdia}f compared with the red H$_2$ contours. As in the previous example, the brightest [\ion{Ne}{2}] emission is clearly associated with the brightest H$_2$ clump spatially. 

Interestingly, despite the very different nature of the starburst region in SQ-A, the PV diagram for SQ-A (Figure~\ref{fig:PVdia}h,i) shows some similarities to the other two examples. Despite there being many H$_2$ components along the line of sight, the [\ion{Ne}{2}] emission, which defines the main HII region (see \citet{Appleton2023}) is shifted slightly (50 \kms) to a lower velocity than the main warm H$_2$ component (this is clear from the white contours in (Figure~\ref{fig:PVdia}i). We had noted previously in the discussion of the spectra, how the peak on the [\ion{Ne}{2}] emission did not quite line up with the peak in the warm H$_2$. The PV diagram shows this clearly, especially for the brighter component of the [\ion{Ne}{2}]. Unlike the cases of the other regions (which tend to show an increase in the velocity of the ionized gas relative to the warm H$_2$), the shift in the [\ion{Ne}{2}] relative to the main H$_2$ peak  in SQ-A is towards lower velocities. This suggests that the ionized gas associated with the starburst component is gas that has been decelerated from a higher velocity by a collision with much lower velocity gas associated with the intruder. This would make sense if the high-velocity gas (6900 \kms) were denser than the intruder gas, as is argued based on CO observations in the cloud-cloud collisional scenario for SQ-A of \citet{Xu2025}.

The spatial coincidence of the intermediate velocity [\ion{Ne}{2}] emission seen in many of the observed regions, and the brightest clumps of H$_2$ in the main intragroup gas suggests that the ionized gas on sub-arcsec scales is formed in close association with the cooler denser molecular gas, and is not just some random projection of intermediate velocity ionized gas spread throughout the galaxy group. This result, of a clear association of decelerated ionized gas with clumps of cool H$_2$ would not have been possible without the remarkable spatial resolution of {\it JWST} MIRI/MRS. 

More insight into the relationship between the different gas phases can be seen in Figure~\ref{fig:fe2spec}. This shows five spectra extracted from the brightest region in Ridge South. As in Figure~\ref{fig:ionspectra}b, we show again the 0-0S(1) H$_2$ line (in red) and the [\ion{Ne}{2}] line (in green). The clear offset in the velocity of the peak emission in both species is very obvious. However, unlike the previous figure we now include three extra spectra. Firstly we show the spectrum of the 0-0S(7) H$_2$ line in orange. Higher-order pure rotational lines of H$_2$ have been shown from previous work on SQ \citep{Cluver2010,Appleton2017} to be associated with warmer, more excited gas that the 0-0S(1) line.  These data show that the peak emission from this higher-excitation 0-0S(7) H$_2$ line (at 6150 \kms) lies in between the highest peak of the cooler H$_2$ line (at 6020 \kms), and the ionized [\ion{Ne}{2}] (at 6200-6250 \kms). This tendency for the warm H$_2$ emission to lie at intermediate velocities was discussed in Guillard et al. (in prep.). We also plot in Figure~\ref{fig:fe2spec} the spectrum of [\ion{Fe}{2}]5.34$\mu$m emission (black line). This gas is only detected in this bright region in Ridge South, and nowhere else except in the high-excitation SQ-A starburst. Similarly, we see that the bulk of the [\ion{Fe}{2}] emitting gas is further shifted to higher velocities, filling in more of the intermediate velocities with stronger emission than the cool H$_2$ as measured in the 0-0S(1) line. We will discuss possible reasons for shift and spread of the warmer H$_2$ and ionized gas to higher velocities in $\S$6. 

Also shown in Figure~\ref{fig:fe2spec} is the CO line profile obtained from the ACA observations with a large beam, which shows that the majority of the very cold gas lies at low velocities corresponding to the red-wing of the 0-0S(1) emission (see $\S$4 for further dicussion). 

\begin{figure}
\includegraphics[width=0.5\textwidth]{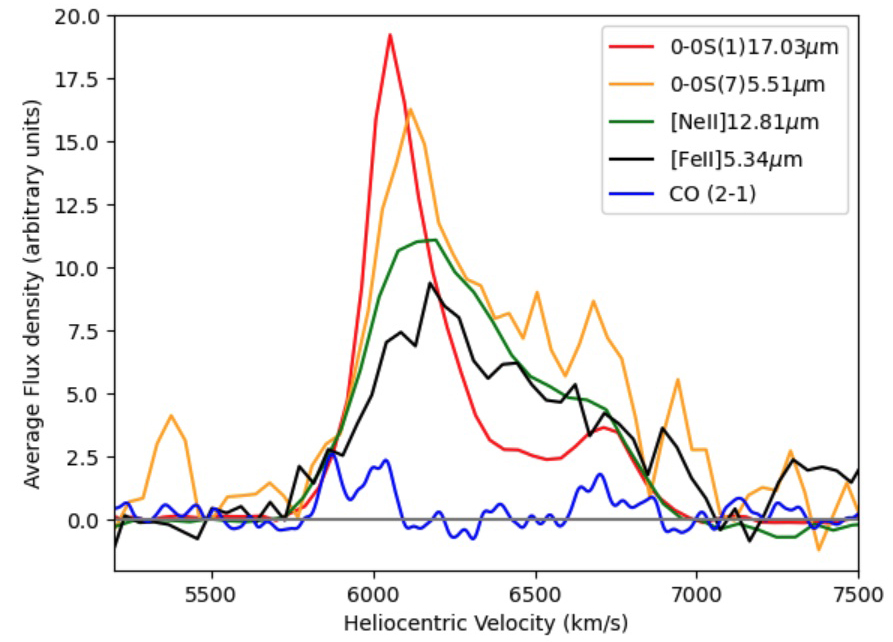}
\caption{Spectra from the brightest compact emission structure in Ridge South (see  Fig.~\ref{fig:ionsmom0}b) : Red line shows the 0-0S(1) H$_2$  line (coolest gas), orange shows 0-0S(7) H$_2$ line (associated with likely hotter gas), green shows [\ion{Ne}{2}] ionized gas, and black shows the [\ion{Fe}{2}]5.34$\mu$m line. The blue spectrum shows faint emission (most significant at the lowest velocities)  in CO (2-1) obtained from an ACA beam centered on the region \citep{Emonts2025}.  Notice the progressive redward shift of the line peak from the cooler low-excitation H$_2$ and CO emission at 6000-6050 \kms~to the warmer H$_2$, ionized Neon gas and [\ion{Fe}{2}] emission which fills the intermediate velocities (6200-6500 \kms).  } 
\label{fig:fe2spec}
\end{figure}

As we have previously noted, SQ-A-South,
which is where the intruder spiral arm probably hit the highest density pre-collision group-wide gas \citep{Xu2025}, stands out in a number of ways 
compared with the other observed regions. As discussed in \cite{Appleton2023}, the area studied is centered on a major cluster of young stars forming roughly an oval shape, the site of a major starburst \citep{Xu1999}. Unlike the other targeted regions, Figure~\ref{fig:MRSpointings}b shows that we detect many different ionized species. In addition to [\ion{Ne}{2}]$\lambda12.81\mu$m and [\ion{Ne}{3}]$\lambda15.56\mu$m, we detect emission from [\ion{Ar}{2}]6.98$\mu$m, [\ion{Ar}{3}]8.99$\mu$m, [\ion{S}{4}]$\lambda10.51\mu$m, [\ion{Fe}{2}]5.34$\mu$m and [\ion{P}{3}]17.89$\mu$m. The Neon lines are very narrow and lie near (but not precisely aligned with) the brightest peak in the warm 0-0S(1) H$_2$ emission. The line-of-sight H$_2$ emission is very complex with many velocity components present, which may explain why the H$_2$ moment map of SQ-A-South is so different from the [\ion{Ne}{2}] map. A full discussion of these lines is beyond the scope of this paper, but in the Appendix, we will argue that the ionized gas in SQ-A-South is photoionized by a young starburst region. This is in agreement with previous studies using optical emission-line diagnostics \citep{Iglesias-Paramo2012,Konstantopoulos2014,Duarte2019,DuartePuertas2021}.

\section{Kinematics of the cold molecular gas}

The CO (2-1) emission, associated with cold molecular gas and observed with the ACA \citep{Emonts2025}, tends to follow the lowest warm H$_2$ velocities along the ridge, and is consistent with gas associated with the intruder galaxy at $\sim$ 6000 \kms. The ACA observations (which use only the 7-m ALMA dishes) are sensitive to only the brightest CO emission on an angular scale of 7-8 arcsecs.
Since the scales of the emission is so different from the {\it JWST} observations, we show only an indicative example in Figure~\ref{fig:fe2spec} of the CO emitting gas from a single ACA beam centered on the same extraction region as the spectra shown in the figure. As can be seen, weak CO emission (representing even colder H$_2$ than the 0-0S(1) emission),  lies close to the lowest velocities (around 5850-6050 \kms) seen in the region, with potentially fainter emission at the velocity of the second peak seen in 0-0S(1) (at 6700 \kms).  Much deeper ALMA 12-m observations at 0.2 arcsec spatial resolution have been made with the same ALMA program and are currently being analyzed, and so we cannot definitively associate the ACA CO emission with the same {\it JWST} region. However, the likely association of CO emission with the extreme ends of the velocity distribution, suggests that colder CO emitting gas is associated with gas that has not been strongly decelerated in the collision.

\section{Excitation of the Ionized Gas}
\label{sec:discussion_models}
In the Appendix, we compare measured line ratios to model predictions based on (i) shocks and on (ii) photoionization from stellar radiation. We base our comparison mostly on the [\ion{Ne}{2}]$\lambda$12.81$\mu$m and [\ion{Ne}{3}]$\lambda$15.56$\mu$m lines which are detected in all regions. In addition, we use the [\ion{S}{4}]$\lambda$10.51$\mu$m which is detected only in SQ-A-South.  Upper limits were calculated for the other regions assuming the same line shape for the [\ion{S}{4}] line as the detected [\ion{Ne}{2}] line profile, and assuming the peak of the undetected emission corresponds to the 3-sigma upper limit for the [\ion{S}{4}] line for that aperture.

As we show in the Appendix, with the exception of SQ-A-South, all the other regions are most compatible with low-velocity shocks as the primary excitation mechanism. 
Despite only having upper limits for [\ion{S}{4}] in these regions, the most compatible shock models suggested pre-shock densities $n$, values of between  $10~<~n~<~100~{\rm cm^{-3}}$, shock velocities of $\sim 100 -200$ \kms, and magnetic field densities of a few $\mu$G. Although very high velocity shocks (V$_{shock}$ $>$ 700 \kms) cannot be ruled out (see Appendix), previous observations in the Ly-$\alpha$ and [\ion{C}{4}] UV lines \citep{Guillard2022} in many of the same regions favored low-velocity shocks to explain the power in these UV lines.   

The one exception regarding the ionized gas excitation is SQ-A-South. The high excitation of the Neon lines, and the firm detection of the [\ion{S}{4}]$\lambda10.51\mu$m line, led us to the conclusion that the ionized gas here is photoionized by a young population of massive stars in a continuous starburst lasting $\sim~3~Myr$. In a  recent study of Stephan's Quintet's compact star clusters, \citet{Aromal2025} support a relatively young age for compact star clusters in or near the SQ-A-South region with ages $< 10^7~yrs$.

\section{Origin of the gas at intermediate velocities on small scales}

To attempt to understand the origin of the gas at intermediate velocities (between that of the intruder galaxy and that of the main group), we review the following observed properties:  i) Almost all the regions studied show the ionized gas lying at intermediate velocities bracketed between the often double-peaked profiles of the 0-0 S(1) H$_2$ line. In some cases the brightest ionized gas is offset by a few hundred \kms~ from those peaks, ii) in many cases the bright ionized gas is spatially coincident with the brightest H$_2$ clumps, suggesting an intimate connection between the ionized gas and the H$_2$ clumps. The gas is not just randomly spread across the region, iii) there is a clear shift away from the velocity extremes in the hotter, more excited gas, compared with the cooler gas. Progressively, higher excitation H$_2$ (e. g. 0-0 S(7) emission), ionized Neon and finally higher excitation [\ion{Fe}{2}] emission are shifted to intermediate velocities between the two velocity extremes, iv) the ionized Neon found within the main ridge and bridge is consistent with excitation in low-velocity shocks, whereas in SQ-A, the site of a major starburst, the ionized gas is more strongly ionized, probably by photoionization.   

An important result of the {\it JWST} observations  is that 
the brightest 0-0~S(1)~H$_2$ emission is associated with either the velocity of the intruder (at radial velocities around 6000 \kms ) or that of the pre-existing tidal filament, at 6700 \kms, and only a relatively small fraction ($\sim$30$\%$) is found at intermediate velocities. Since most of the observed warm H$_2$ emission is emitted from the velocity extremes rather than from the intermediate velocities, there must have existed pre-existing dense molecular clumps which have largely survived the initial shock, and maintained their momentum.  One idea put forward in Guillard et al. (in prep.), is that the majority of the warm H$_2$ emission is heated close to the cold dense clump (a result supported by the detection of strong CO emission at the velocity of the intruder along the main north/south filament \citep{Emonts2025}) in a mixing layer where much of the heating and energy dissipation is taking place. 
\subsection{Excitation-dependent decelerations of the gas relative to the intruder's velocity}
How then do we explain the kinematic behavior of gas in Figure~\ref{fig:fe2spec}, which clearly shows that gas of increasing excitation is being decelerated away from the intruder velocity (around 5800-6000 \kms) to higher and higher velocities?  In a very simplistic way, it is easy to understand why gas of differing initial densities may be accelerated more (or less) by a hot flow. For example, by examining models used to explain ionized gas flows in supernova blast waves, which represent a similar situation to that of a strong shock wave passing over pre-existing gas clumps in SQ, \citet{McKee1975} considered a blast wave traveling with a velocity $v_w$ in a medium of density $\rho_0$ encountering a denser nearly spherical cloud of density $\rho_c$. For an ideal gas, a shock wave would be driven into the cloud with a velocity v$_c$, such that  $v_c^2 \sim \chi v_w^2$, where $ \chi = \rho_0 / \rho_c$, the density contrast between the hot medium from the blast wave and the dense clump. If we assume for SQ, $v_w$ = 500 \kms and  $\rho_0 = 10^{-2} cm^{-3}$ (reasonable given the intruder's impact velocity and the density of the presumed diffuse group medium), then shock waves would travel through the cloud over a range of velocities from 50 \kms (for $\rho_c = 1~cm^{-3}$) to 2 \kms~for very dense clouds ($\rho_c = 10^3 cm^{-3}$). Lower density clouds would tend be quickly overrun by the shock, accelerated and ionized, based on their so-called cloud crushing timescale, which is the time it takes the cloud shock to cross the cloud,  $t_{cc} = r_c/v_s$, or $\chi^{1/2} r_c/v_b$, where $r_c$ is the characteristic radius of the cloud. For an initially low density cloud caught up in the flow,  ($\rho = 1~cm^{-3}$) and a cloud size of say 10~pc, $t_{cc} = 0.2 Myr$. A much higher density cloud (say a molecular cloud) with $\rho_c = 10^3 cm^{-3}$, the time to crush the cloud would be longer, 6 Myr, implying such clouds might survive over the timescale of the galaxy collisions in SQ (believed to be approximately 10$^7$ yrs). Of course these cloud crushing times may be shorter or longer depending on the assumed shock speed and cloud size. For example, previous shock modeling of the H$_2$-excitation by \citet{Appleton2017} suggested that the slowest C-shocks needed to explain the gas excitation was 5 \kms, and so the cloud crushing time would be 2 Myr for a 10pc cloud, or 20 Myr for a 100 pc cloud. However, faster shocks were also implied, which would further reduce $t_{cc}$.  

It is therefore tempting to associate low density clouds, which would be quickly destroyed and accelerated to the velocity of the flow, with the [\ion{Ne}{2}] ionized gas, and the initially denser clouds with the warm H$_2$ emission. Denser clouds would be harder to accelerate (through a constant ram pressure from the wind which goes as $\rho_0 v_w^2$), and therefore having clouds of different initial densities could explain the differential acceleration of the different gas components seen in Figure~\ref{fig:fe2spec}). In this picture all that is needed to explain the differential acceleration (deceleration relative to the intruder) of the various components seen in Figure~\ref{fig:fe2spec} is to have a hot flow of gas from the group-wide medium encounter pre-existing material of different initial densities in the intruder galaxy,  leading to differential acceleration of the different clumps. In this picture it is not necessary for the low and high density pre-collision medium to be spatially correlated. For example, it would not explain point ii) at the beginning of this section. 

As we showed in Figure~\ref{fig:fe2spec}, the [\ion{Fe}{2}] emission is strongly shifted in velocity away from the lower velocity peak of the cooler H$_2$ gas, and is spread over a broader range of velocities. 
[\ion{Fe}{2}] emission is often associated with the shattering of iron atoms on grain surfaces in fast shocks by sputtering processes as pre-shock grains pass into the rapidly decelerated post shock gas \citep[e. g. ][]{Jones1994,Draine1995}). \citet{Jones1994} showed that the shocks with velocities greater than 100-150 \kms will result is a large depletion of the Iron grains (50-80 $\%$ destruction) into the gas phase. So the gas associated with the [\ion{Fe}{2}] line may be more strongly shocked than the Neon emission (see Appendix). This again may be the result of a shocks passing into much lower density material where is shock speed would be higher. 

\subsection{Acceleration and survival of the molecular gas}
While the above picture may be helpful at explaining the ionized gas acceleration in Figure~\ref{fig:fe2spec}, there are complications when considering cold molecular gas. Models of molecular gas clouds in a hot flow show that it is hard to accelerate denser cold clouds without them being destroyed before they can be accelerated. Indeed this has always been the underlying issue about how molecular gas can survive in hot galactic halos \citep{Faucher2023}.  For example \citet{Girichidis2021} showed that the timescale needed to accelerate a dense molecular cloud into a fast flow, $t_{drag} = \chi^{1/2} t_{cc}$, which implies for realistic values of $\chi >> 1$, the clouds would be destroyed by a variety of hydrodynamic instabilities before they can be accelerated. This is partly because it is not only the cloud crossing time that is important, but also the flow-time of the hot wind over the cloud which is much shorter. 

One solution to the problem comes from the clouds growing if the colder gas can mix with the hot wind, especially when combined with strong radiative cooling \citep{Abruzzo2022,Gronke2018,Gronke2022}.  
In recent numerical gas dynamical models, \citet{Farber2023} show that molecular clouds with temperatures of 300 K can survive a fast flow under certain conditions. An important parameter in their models is the interplay between the gas cooling rate, $t_{cool}$ and $t_{cc}$. Radiative cooling is strong through the warm H$_2$ lines in this case, and $t_{cool}$ is very short \citep{Appleton2017}, of the order of a few thousand years.  In \citet{Farber2023}, clouds larger than a critical size will shatter into a mist of smaller clouds which can interact and mix with the hot flowing gas. By shattering, they can cool down and remain molecular. This may explain the hotter molecular gas (0-0S(7) H$_2$) making up 30$\%$ of the gas that has been decelerated in Figure~\ref{fig:fe2spec}. On the other hand, the same models showed that clouds smaller than a critical radius, will also shatter, but fail to cool down to molecular temperatures. Instead these clouds remain hot and of low density. Although such models are very uncertain, we mention them because they can provide another explanation for why the ionized gas is easily entrained in the hot flow, and would end up with larger offsets from their initial pre-shock velocities (as in the [\ion{Ne}{2}] and [\ion{Fe}{2}] profile shapes in Fig.~\ref{fig:fe2spec}).  

It is possible that both the simple ram-pressure blast wave, and the mixing layer pictures for the multi-phase gas are occurring together across the main shock-front of the Quintet. The denser cold molecular seed regions in the pre-existing gas could generate mixing layers over their surfaces which protect and prolong the existence of the dense clump as it experiences a hot wind (through mixing and rapid cooling with the hot medium). Lower-density regions, ablated from the clump, would be preferentially quickly accelerated, whereas the higher density molecular clumps (part of the outer regions of the mixing layer) would be torn off and be more slowly accelerated. This is another explanation for the behavior of the spectrum shown in Figure~\ref{fig:fe2spec}.  This picture, which combines both simple ram pressure acceleration with ablated dense clumps might explain the correlation between the denser H$_2$ clumps and the isolated but accelerated ionized gas, as well as the obvious head-tail morphologies seen in many of the clumps along the main shock (see discussion by \citet{Appleton2023}. 

While some of the warm H$_2$ detected by {\it JWST} may represent transiently heated low-density gas which is shock-heated quickly to an ionized medium, the question of the survivability of the warm H$_2$ emission in the hot halo of the Quintet requires a more complex solution.  Dense preexisting cold molecular clouds in the intruder galaxy can experience a strong hot wind created by the collision (Guillard et al. in prep.), causing the development of a turbulence close to the cold clump. The lifetime of the clumps is not well determined, either theoretically (models which deal with supersonic turbulence in such situations are still in their infancy) or observationally.  If we interpret the length-scale of warm H$_2$ structures along the main filament as typically 5 kpc (and in one case 15 kpc) from the MIRI images of \citet{Papadopoulos2008} as long tails resulting from the ablation of dense clumps in the hot wind (of say 500 \kms),  the cloud lifetimes would be 10-30 Myr. This is significantly longer than the cooling time of the warm H$_2$ of a few thousand years \citep{Appleton2017}. This disparity in timescales supports the idea of turbulent mixing layers as a way to extend the life of a cloud through mixing and rapid cooling of the gas. Such potential mixing layers would be the site of the shock-heating of the molecular gas creating the strong warm H$_2$ emission observed.

\subsection{Comparison with molecular gas in the tails of ram-pressure-stripped galaxies}

As mentioned in the introduction, ram-pressure stripped galaxies in clusters often exhibit tails which contain ionized and molecular gas (sometimes called Jellyfish galaxies; see for example \citealt{Fossati2016,Jachym2019,Moretti2018}). This material is believed to have been stripped out of gas-rich galaxies through the hydrodynamical effects of the hot intracluster medium on the galaxy as it traverses the cluster. There is some evidence that the compression caused by the interaction can convert HI into H$_2$ within the galaxies , leading to stripping of that gas into a tail, some of which might lead to {\it situ} formation of both the cold gas and HII regions \citep{Fossati2016,Moretti2020,Moretti2023}. 

One could argue that the mass conversion of the pre-collisional HI-filament into molecular gas by the powerful shock-wave caused by the intruder in SQ, is an extreme example of this phenomenon (see \citet{Cluver2010} for more discussion). It is hard to make detailed comparisons between the warm H$_2$ discussed in this paper with the published HI observations, because of the huge difference in spatial resolution between the sub-arcsec MIRI observations, and the highest resolution HI observations (8 x 7 arcsec$^2$--see \citealt{Xu1999}). A more complete discussion of the HI emission relative to the comparable low-resolution Atacama Compact Array CO observations is described in detail in \citet{Emonts2025}.  Until recently, it was believed that little or no HI had survived the large-scale shock wave driven into the pre-existing HI filament, since no HI was found along the main molecular filament \citep{Emonts2025}. However, faint HI corresponding to the main molecular filament (obtained at a spatial resolution of 17 x 14 arcsec$^2$) is now seen in the very deep HI observations of \citet{Rajpurohit2026a}. This suggests that at least some HI survived the main high-speed shock in SQ.   

The question of the survival of molecular gas to large distances from the parent galaxy in the Jellyfish systems may provide some parallels with SQ. For example, \citet{Sparre2020} have presented models of the development of mixing layers on the surface of clouds in the tails of Jellyfish galaxies to help explain their survival. Such models, though developed for warmer clouds (T$\sim$10$^4$ K), develop long tails, which is similar to those seen in SQ. 

The gas in the tails may also survive for longer if they are supported by magnetic fields. For example, magnetic draping of clouds has been invoked to help to explain the length and longevity of Jellyfish tails.  \citet{Sparre2020} discuss how magnetic fields can stabilize a gas cloud caught up in a strong wind. The models show that the compression of magnetic fields up-stream of a cloud, and the magnetic draping of dense gas clumps in the tail, can stabilized the system against hydrodynamic instabilities. Also, because of adiabatic compression and the effects of shear in the tails, the models predict an alignment of the B-field with the direction of the tails. This appears born out in some cases, for example, in the Jellyfish galaxy JO206 \citep{Muller2021}. Here, unusually high fractional polarization in the tails of $>$ 50$\%$ are found, along with close alignment of the B-vector to the tail direction.   

Something like this is possible, but to a lesser degree, in SQ, where the majority of the synchrotron emission from SQ is well explained by the shock acceleration of cosmic ray electrons, as the intruder galaxy collided with the pre-existing HI structure \citep{Rajpurohit2026b}. The southern part of the main N/S filament shows 5$\%$ polarization and approximate alignment of the B-vector with the overall filament \citep{Nikiel2013}. However, much of the rest of the structure is less clearly aligned with the structures, and so the dominance of magnetic effects in SQ may be less clear than in the Jellyfish systems. 

\section{Conclusions}

This paper is one of a series of papers presenting observations of the multi-phase intergalactic medium in Stephan's Quintet using the {\it JWST}  Mid-Infrared Instrument (MIRI) Medium Resolution Spectrometer (MRS). The MRS covers the wavelength range from 4.9 to 27.9$\mu$m, at a spatial scale ranging between 0.2 to 0.27 arcsec/pix. Future papers will combine new high-resolution CO observations with the {\it JWST} observations to better describe the spatio-kinematics behavior of the emission.  In the current paper, based on a comparison between the distribution and kinematics of [\ion{Ne}{2}]$\lambda12.81\mu$m and [\ion{Ne}{3}]$\lambda15.56\mu$m,  and the pure rotational lines of warm molecular hydrogen (especially the strong 0-0S(1)$\lambda17.03\mu$m line) in six targeted regions in the Quintet, as well as the line strengths of ionized [\ion{Ne}{2}], [\ion{Ne}{3}] and [\ion{S}{4}]$\lambda10.51\mu$m,  we find that: 
\begin{itemize}
\item Representative spectra associated with five of the six pointings show a common behavior. In most cases the spectrum of warm molecular hydrogen (0-0S(1)$\lambda$17$\mu$m line) show  double peaked line profile with most of the warm H$_2$ piled up at either the approximate velocity of the intruder galaxy NGC 7318b ($\sim$~6000 \kms) or at the expected velocity of the preexisting tidal filament (6700 \kms).  A small fraction of the warm H$_2$ is observed between the two extreme velocities. In contrast, the majority of the emission from ionized Neon is shifted away from the main peaks or lies completely between them. In Ridge North, some of the [\ion{Ne}{2}] emission peaks at the lowest velocities, although faint emission is also seen between the twin peaks of the warm $H_2$. The observed behavior suggests that, unlike the majority of the colder H$_2$ (CO emitting and the 0-0S(1) rotational emission) which has preserved its pre-collisional momentum,  the ionized gas, relative to the intruder's velocity, has been strongly decelerated in the collision between the intruder galaxy and the intragroup medium. One exception is the region associated with SQ-A, a known intergalactic starburst system, where the H$_2$ line profiles show up to five separate components. Here the ionized gas lines are slightly offset bluewards of the brightest spectral H$_2$ component extracted at the position of the starburst.   

\item On the arcsec or subarcsec scale (scales $<$ 450 pc) in many cases the brightest ionized gas clumps are strongly correlated with the brightest warm H$_2$ clumps. This is consistent with the idea that cold, dense, pre-existing molecular gas clumps have interacted with the opposing flow of gas in the collision, stripping off and shock-heating lower density gas which is quickly accelerated and ionized. An exception is Ridge North, which shows significant [\ion{Ne}{2}] emission at the velocity of the intruder, whereas much of the warm H$_2$ is seen at much higher velocities.

\item In the Ridge South region at the position of the most compact knot seen in both H$_2$ and Neon emission, we show an interesting result. The more the gas is decelerated away from the intruder's velocity, the more it is progressively heated and increasingly more strongly ionized. Cold CO-line emitting, and low-excitation pure-rotational H$_2$ emission (in the 0-0S(1) line) is less shifted in velocity than higher excitation warmer H$_2$ (0-0S(7) line dominated), which in turn is less shifted than ionized Neon (originating in low-velocity shocks). Finally, even more extensively shifted and decelerated, is the [\ion{Fe}{2}] emitting gas which is only detected in this one region. Shock velocities in excess of 100 \kms \citep{Jones1994} are required to significantly deplete Iron grains from the solid to the gas phase.  These excitation-dependent decelerations are likely a result of gas of different initial pre-shock cloud densities being acted upon as they move against a hot wind driven by the collision of the intruder with the IGM of the group. 
Lower-density pre-existing clouds, in the intruder galaxy, would experience faster cloud-shocks (resulting in more heating and ionization), than much denser molecular clouds \citep{McKee1975}, as they encounter an opposing gas flow originating in the intragroup medium. Lower density clouds would also be easier to accelerate via ram-pressure.    

\item As discussed in Guillard et al. (in prep.), dense cold clouds entrained in a wind may not simply develop a single low-velocity shock structure (see \citealt{McKee1975}). Instead, they can develop turbulence at their surfaces, creating  mixing layers that can prolong the lifetime of the molecular gas and recycle hot gas into the cold phase. Gas stripped from these clumps can break away and be accelerated. Lower density gas is easier to accelerate, and can quickly become ionized, naturally explaining the spatial correlation between warm H$_2$ and ionized gas on sub-arcsec/arcsec scales seen here.    

\item Using the line ratios of [\ion{Ne}{3}]/[\ion{Ne}{2}], and upper limits to the emission from [\ion{S}{4}]$\lambda10.51\mu$m, we are able to show that the majority of the ionized Neon emission is consistent with either low-velocity shocks ($< 250$ \kms), or very high velocity shocks.  Although very high velocity shocks ($\geq 800$ \kms) cannot be ruled out by these current data, previous HST UV spectroscopy (especially upper limits to the  detection of [\ion{O}{6}]$\lambda$1032$\AA$ emission which is expected to be strong in fast shocks \citep{Guillard2022}) strongly favors lower velocity shock interpretation. 
The only regions that shows very strong ionized gas excitation is SQ-A-South, where \citet{Xu2025} has argued for an unusually high speed collision between gas in the group. Here, the strongly detected [\ion{S}{4}] emission, combined with the high [NeIII/NeII] ratio is incompatible with most shock models, but is easily explained by photoionization from a recent starburst with an age of at least 3 Myr. This is in agreement with a recent study of compact star clusters in the same region \citep{Aromal2025}. 

\end{itemize}
\vspace{-0.425 in}
Future observations with at higher spatial resolution in the CO line, combined with current and future {\it JWST} observations will help to better determine whether mixing layers surrounding cold dense clouds are a viable way to extend the life of molecular gas in the group-wide halo of SQ. 

\begin{acknowledgements}

 This work is based, in part, on observations made with the NASA/ESA/CSA James Webb Space Telescope. These data were obtained from the Mikulski Archive for Space Telescopes at the Space Telescope Science Institute, which is operated by the Association of Universities for Research in Astronomy, Inc., under NASA contract NAS 5-03127 for JWST. The authors acknowledge support under NASA Guest Observer grant JWST-GO-03445 001-A. The National Radio Astronomy Observatory and Green Bank Observatory are facilities of the U.S. National Science Foundation operated under cooperative agreement by Associated Universities, Inc. The project is also supported by the Sorbonne University, the Institut Universitaire de France, the Centre National d'Etudes Spatiales (CNES), the `Programme National de Cosmologie and Galaxies' (PNCG) and the `Physique Chimie du Milieu Interstellaire' (PCMI) programs of CNRS/INSU, with INC/INP co-funded by CEA and CNES,  for there financial supports. UL acknowledges support from the project PID2023-150178NB-100, as well as from the  Consejer\'ia de Universidad, Investigaci\'on e Innovaci\'on and Gobierno de Espa\~na and Uni\'on Europea - NextGenerationEU through grant AST22$\_$4.4, financed by MCIN/AEI/10.13039/501100011033, and FQM108, financed by the Junta de Andaluc\'ia (Spain). Datasets: The {\it JWST} data used in this paper can be found in MAST: \dataset[10.17909/109z-0224]{http://dx.doi.org/10.17909/109z-0224}.

\end{acknowledgements}

\newpage

\facility{{\it JWST}/MIRI-MRS}
\software{scipy \citep{scipy2001}, astropy \citep{astropy2013,astropy2018,astropy2022}, SAOimage/DS9 \citep{Joye2003}, Jdaviz/Cubeviz \citep{jdadf2024}, SPECTRACUBE \citep{Ginsburg2015}}

\appendix

\section{Ionized Gas Excitation}
\label{sec:app_excite}

\begin{figure*}
\figurenum{A1}
\includegraphics[width=0.5\textwidth]{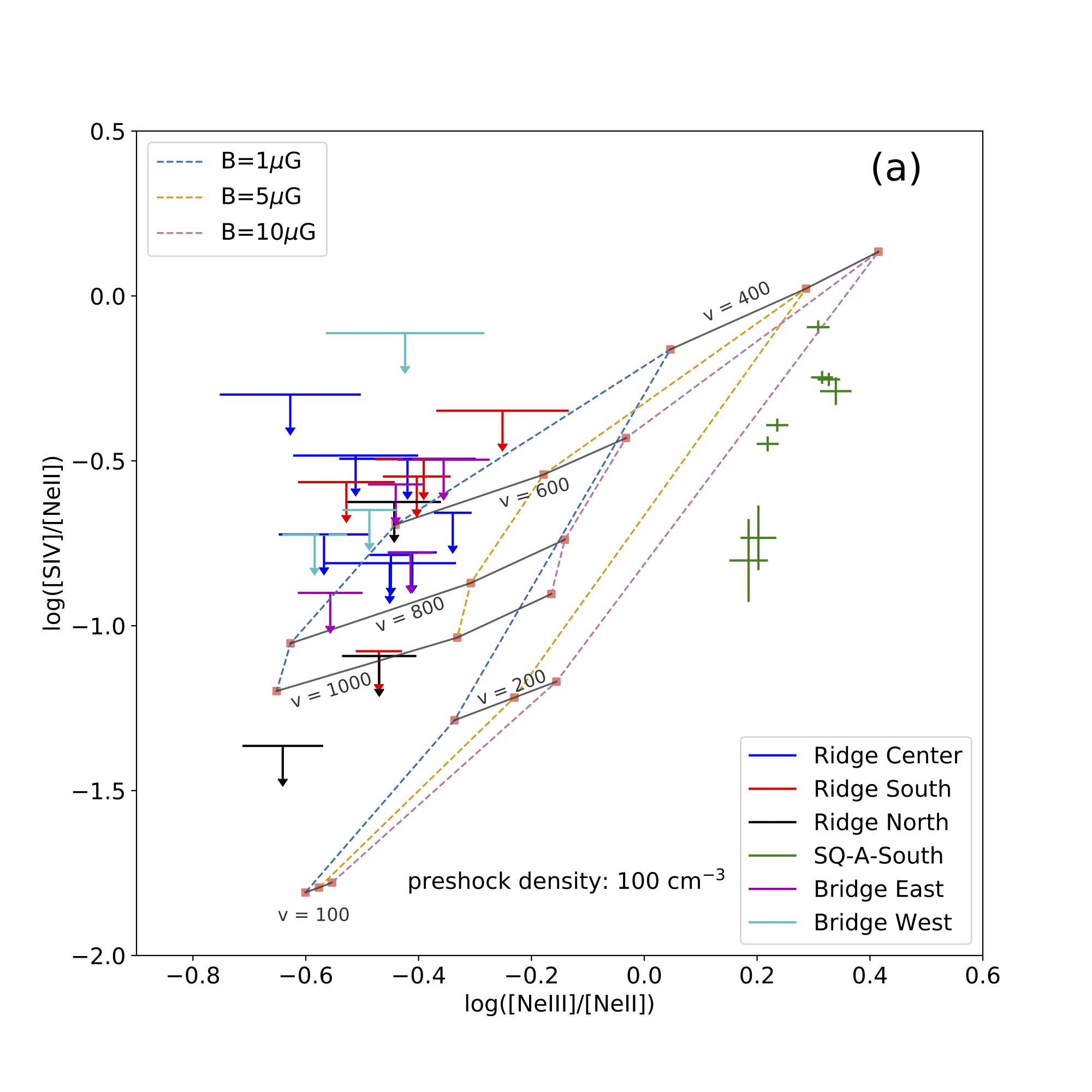}
\includegraphics[width=0.5\textwidth]{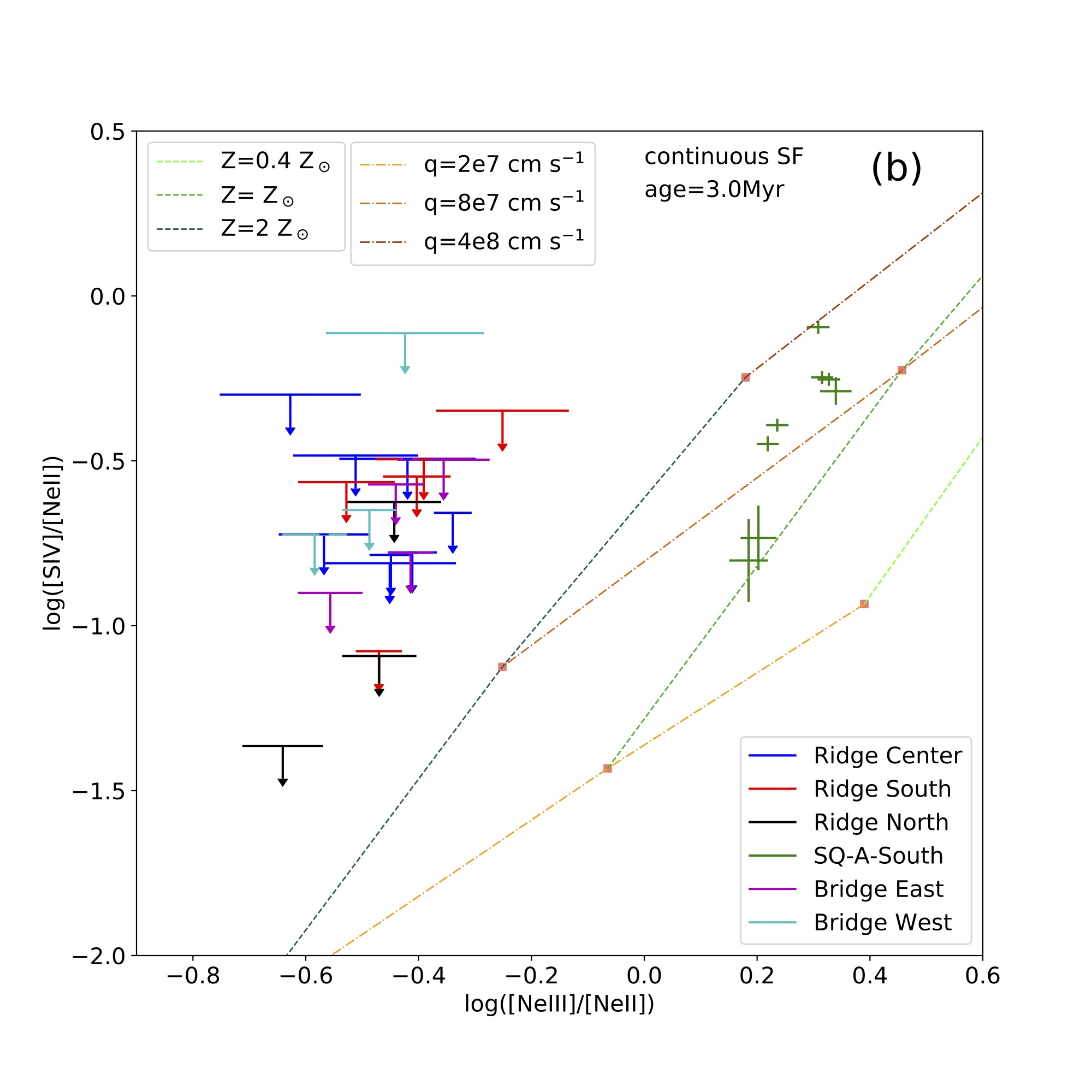}

\caption{a) Comparison of the observed [\ion{S}{4}]$\lambda$10.51$\mu$m/[\ion{Ne}{2}]$\lambda$12.81$\mu$m vs. [\ion{Ne}{3}]$\lambda$15.56$\mu$m/[\ion{Ne}{2}]$\lambda$12.81$\mu$m line ratios to shock  models based on the MAPPINGS III model \citep{allen2008} for regions where the Neon lines have been observed with a S/N $>$ 3 within the different observing regions. The prediction is for a shock model with a precursor for absolute Solar metallicity (Z = 0.02) and a preshock gas density of $n = 100 {\rm cm^{-3}}$. The green and brown lines give the values for different magnetic field strengths and the green lines for different shock velocities,  b) Comparison of the same observed line ratios as (a) but using the grid of \citet{Inami2013} (see text). The model predictions are for a continuous starburst of age 3~Myr and an electron density of $n_{\rm e} = 100$ cm$^{-3}$. Results are shown for different metallicities, Z, and ionization parameters, q.
 The green and brown lines give the values for different metallicities and ionization parameters. 
 }
\label{fig:ionmodels}
\end{figure*}

In this Appendix we use the [\ion{S}{4}]$\lambda$10.51$\mu$m, [\ion{Ne}{2}]$\lambda$12.81$\mu$m and [\ion{Ne}{3}]$\lambda$15.56$\mu$m line ratios to investigate the excitation of the ionized gas in the six target regions. As discussed in \S 2, emission line ratios were measured for multiple regions covering the six targeted MIRI fields (typically 3 to 10 apertures in each targeted observation). The regions were chosen to capture obvious features in each cube. We will describe the process in more detail in a future paper. 

For the shock modeling, we use the library from the model MAPPINGS III \citep{allen2008} to predict the expected line ratios of our measured lines. This library considers shocks with velocities between 100 and 1000 \kms\ for different magnetic parameters, metallicities and preshock densities. We use models for an assumed Solar metallicity, which is in general agreement with the oxygen abundance measurements along most of the main filament measured by \citep{DuartePuertas2021}, although metallicities as low as half-Solar are found in some regions, but these are mainly outside our areas of interest. 

The model library of \citet{Inami2013} was also used to predict Neon and Sulphur line ratios for the possible case of photoionzation by young stars. \citet{Inami2013} used the photoionization models from \citet{levesque2010} based on Starburst99 and the MAPPINGS III code. In these models, the Starburst99 stellar population synthesis code \citep{leitherer1999,vazquez2005}  was used to generate stellar spectral energy distributions (SEDs), which are then used as inputs to the Mappings III photoionization code \citep{binette1985, sutherland1993, groves2004} to predict the resulting nebular emission. The models are calculated for a range of stellar metallicities (which are assumed to be the same for the stars and the gas), a Salpeter Initial Mass Function, two mean electron densities ($n_{\rm e} = 10$ and 100 cm$^{-3}$) and ionization parameters $q$ ranging from $1 \times 10^7$ to $4\times 10^8$ cm\,s$^{-1}$. 

Figures~\ref{fig:ionmodels}a and b present the [\ion{S}{4}]/[NII] versus [\ion{Ne}{3}]/[\ion{Ne}{2}] line ratios for the six observed regions for shock and photoinization. There is a striking difference between SQ-A-South (green points) and the other regions. Unlike the other regions, the line fluxes in SQ-A-South are much stronger and  [\ion{S}{4}] is easily detected, whereas only upper limits are obtained elsewhere. Furthermore, SQ-A-South is the only region where [\ion{Ne}{3}] is brighter than [\ion{Ne}{2}] (log~[\ion{Ne}{3}]/[\ion{Ne}{2}] $\sim 0.2-0.5$), while it ranges between -0.8 and -0.25 in the other regions. 

Figures~\ref{fig:ionmodels}a presents the line ratios in comparison with one example of a shock model grid (pre-shock density $n  = 100~{\rm cm^{-3}}$ and a range of shock velocities and magnetic field strengths) superimposed. Although we explored a range of preshock densities $n$, values of between  $10~<~n~<~100~{\rm cm^{-3}}$, and shock velocities of $\sim 100 -200$ \kms, are the most compatible with the observed line ratios for all the regions (except SQ-A-South). As seen in Figure~\ref{fig:ionmodels}a, as the shock velocity is increased, the models predict an increase in [\ion{S}{4}]/[\ion{Ne}{2}] ratio up to a shock velocity of 500 km/s. At higher velocities the ratio begins to drop again. Thus, very high velocity shocks $>700$ \kms\ are potentially compatible with the [\ion{S}{4}]/[\ion{Ne}{2}] upper limits and low values of [\ion{Ne}{3}]/[\ion{Ne}{2}] ratio.  However, previous observations in the UV by \citet{Guillard2022} strongly support much lower-velocity shocks.  Models showed that above $\sim$200 \kms, shocks should generate rapidly rising emission from [OVI]$\lambda1032\AA$, whereas this line is not detected. Below this threshold, the observed emission line ratios of Ly-$\alpha$/H-$\beta$ and [\ion{C}{4}]/H-$\beta$ are consistent with much lower velocity shocks. Taken together with the current [\ion{Ne}{2}]/[\ion{S}{4}] upper limits, the results favor lower velocity shocks.  

It is obvious from Figure~\ref{fig:ionmodels}a,  that SQ-A-South (green symbols) has observed line ratios that are not compatible with pure shock models. Although not shown here, there are MAPPINGS models of a strong shocks that include a UV-generating precursor, that could explain the high [\ion{Ne}{3}]/[\ion{Ne}{2}] ratio in SQ-A-South. These possibilities will be presented in a separate paper. However, given that SQ-A is known to exhibit recent star formation, we instead explore photoionization as a more likely alternative explanation for the high excitation of the lines in SQ-A. The predicted photoionization model line ratios vary depending on whether the star formation is instantaneous or continuous.  The model predictions for an instantaneous starburst (not shown here) change very quickly in time. Again assuming solar metallicity (Z = 0.02), the lines ratios observed in SQ-A-South can be explained by a starburst of age $\sim 1.5$~Myr, independent of the electron density, but rapidly diverge from the observations for ages $>~2$~Myr. Continuous star formation is expected to be a more realistic model because the high SFR generally lasts for a few Myr.  Good agreement with the line ratios in SQ-A-South can be achieved for $t \gtrsim$ 2.5~Myr, independent of the electron density. After this time, the predictions of the model are practically constant, so that we cannot exactly determine the age of the starburst from our line ratios other than concluding that it is older than $\sim$ 2.5~Myr. A continuous star formation model at 3~Myr is shown in Figure~\ref{fig:ionmodels}b.


\bibliographystyle{aasjournal}
\bibliography{paper.bib}

\end{document}